\documentclass{appolb}
\usepackage{epsfig}
\usepackage{scalefnt}

\begin{document}

\title{Resonances $f_0(1370)$ and $f_0(1710)$ as scalar $\bar q q$ states
from an $N_f=3$ Sigma Model with Vectors and Axial-Vectors\thanks{%
Presented at the Excited QCD 2011 Workshop, Les Houches (France), 20-25
February 2011}}
\author{Denis Parganlija 
\address{Institute for Theoretical Physics, Goethe University,
Max-von-Laue-Str.\ 1, D--60438 Frankfurt am Main, Germany 
 \\       
parganlija@th.physik.uni-frankfurt.de} }
\maketitle

\begin{abstract}
Features of the non-strange $\bar n n$ and strange $\bar s s$ scalar mesons
are investigated in the Extended Linear Sigma Model (eLSM) with $N_f=3$ and
vector and axial-vector mesons. Our model contains a pure non-strange and a
pure strange scalar state; implementing the mixing of the two states
originates two new states, a predominantly non-strange and a predominantly
strange one. We investigate the possibility to assign the two mixed states
to experimentally well-established resonances. To this end, we calculate the
masses and the two-pion decay widths of the mixed states and compare them
with experimental data. The predominantly non-strange state is found to be
consistent with the resonance $f_0(1370)$ and the predominantly strange
state with the resonance $f_0(1710)$.
\end{abstract}


\PACS{12.39.Fe, 12.40.Yx, 14.40.Be, 14.40.Df}

\section{Introduction}

Experimental data \cite{PDG} show an abundance of scalar meson resonances
both in the non-strange and in the strange sectors. In particular the kaons
and other mesons containing strange quarks are expected to play an important
role in vacuum phenomenology as well as in the restoration of the $%
U(N_{f})_{L}\times U(N_{f})_{R}$ chiral symmetry \cite{Reference1}, a
feature of the Quantum Chromodynamics (QCD) broken in vacuum spontaneously 
\cite{Goldstone} by the quark condensate and explicitly by non-vanishing
quark masses ($N_{f}$ denotes the number of quark flavours). Meson
phenomenology in the non-strange and strange sectors has been considered in
various sigma model approaches (see Ref.\ \cite{Reference1} and references
therein). In this paper, we present an Extended Linear Sigma Model (eLSM 
\cite{Krakow,Madrid}) where these approaches are generalised to contain
scalar, pseudoscalar, vector and axial-vector mesons both in the non-strange
and strange sectors. In particular, we devote attention to the structure of
scalar mesons $f_{0}(1370)$ and $f_{0}(1710)$. In Ref.\ \cite{Paper1}, the
resonance $f_{0}(1370)$ was found to be predominantly of $\bar{q}q$
nature [thus disfavouring the interpretation of the scalar state $f_{0}(600)$
as a $\bar{q}q$ state]. However, the model of Ref.\ \cite{Paper1}
contained no strange mesons and in this paper we address the question
whether the conclusion of Ref.\ \cite{Paper1} regarding $f_{0}(1370)$\ as a predominantly $%
\bar{q}q$ state also holds in a more general $U(3)_{L}\times U(3)_{R}$
approach that simultaneously allows for a statement regarding the structure of $f_{0}(1710)$
(which we find to be predominantly of $\bar{s}s$ nature).

The paper is organised as follows. In Sec.\ 2 we present the model
Lagrangian, the results are discuss in Sec.\ 3 and in Sec.\ 4 we provide a
summary and outlook of further work.

\section{The Model}

The Lagrangian of the Extended Linear Sigma Model with $U(3)_{L}\times
U(3)_{R}$ symmetry reads \cite{Reference1,Krakow,Madrid}: 
\begin{eqnarray}
\mbox{\fontsize{10}{9}\selectfont $ \lefteqn{\mathcal{L}=\mathrm{Tr}[(D^{\mu }\Phi )^{\dagger }(D^{\mu }\Phi)]-m_{0}^{2}\mathrm{Tr}(\Phi ^{\dagger }\Phi )-\lambda _{1}[\mathrm{Tr}(\Phi^{\dagger }\Phi )]^{2}-\lambda _{2}\mathrm{Tr}(\Phi ^{\dagger }\Phi )^{2}} $} \nonumber \\
& & \mbox{\fontsize{10}{9}\selectfont $ - \, \frac{1}{4}\mathrm{Tr}[(L^{\mu \nu })^{2}+(R^{\mu \nu })^{2}]+\mathrm{Tr}\left[ \left( \frac{m_{1}^{2}}{2}+\Delta \right) (L^{\mu })^{2}+(R^{\mu})^{2}\right] +\mathrm{Tr}[H(\Phi +\Phi ^{\dagger })] $} \nonumber \\
&&  \mbox{\fontsize{10}{9}\selectfont $ + \, c_{1}[(\det \Phi +\det \Phi ^{\dagger })^{2}-4\det (\Phi \Phi ^{\dagger})]-2ig_{2}(\mathrm{Tr}\{L_{\mu \nu }[L^{\mu },L^{\nu }]\}+\mathrm{Tr}\{R_{\mu \nu }[R^{\mu },R^{\nu }]\}) $}   \nonumber \\
&&  \mbox{\fontsize{10}{9}\selectfont $ +\, \frac{h_{1}}{2}\mathrm{Tr}(\Phi ^{\dagger }\Phi )\mathrm{Tr}[(L^{\mu})^{2}+(R^{\mu })^{2}]+h_{2}\mathrm{Tr}[(\Phi R^{\mu })^{2}+(L^{\mu }\Phi)^{2}]+2h_{3}\mathrm{Tr}(\Phi R_{\mu }\Phi ^{\dagger }L^{\mu }) $} \nonumber \\
\label{Lagrangian}
\end{eqnarray}
where
\begin{equation}
\scalefont{0.81}\Phi =\frac{1}{\sqrt{2}}\left( 
\begin{array}{ccc}
\frac{(\sigma _{N}+a_{0}^{0})+i(\eta _{N}+\pi ^{0})}{\sqrt{2}} & 
a_{0}^{+}+i\pi ^{+} & K_{S}^{+}+iK^{+} \\ 
a_{0}^{-}+i\pi ^{-} & \frac{(\sigma _{N}-a_{0}^{0})+i(\eta _{N}-\pi ^{0})}{%
\sqrt{2}} & K_{S}^{0}+iK^{0} \\ 
K_{S}^{-}+iK^{-} & {\bar{K}_{S}^{0}}+i{\bar{K}^{0}} & \sigma _{S}+i\eta _{S}%
\end{array}%
\right)  \normalsize \label{Phi}
\end{equation}%
%
%
%
%
%
is a matrix containing the scalar and pseudoscalar degrees of freedom, $%
L^{\mu }=V^{\mu }+A^{\mu }$ and $R^{\mu }=V^{\mu }-A^{\mu }$ are,
respectively, the left-handed and the right-handed matrices containing
vector and axial-vector degrees of freedom\ with 
\begin{equation}
\scalefont{0.81}V^{\mu }=\frac{1}{\sqrt{2}}\left( 
\begin{array}{ccc}
\frac{\omega _{N}+\rho ^{0}}{\sqrt{2}} & \rho ^{+} & K^{\star +} \\ 
\rho ^{-} & \frac{\omega _{N}-\rho ^{0}}{\sqrt{2}} & K^{\star 0} \\ 
K^{\star -} & {\bar{K}}^{\star 0} & \omega _{S}%
\end{array}%
\right) ^{\mu }{\normalsize ,}\;\scalefont{0.81}A^{\mu }=\frac{1}{\sqrt{2}}%
\left( 
\begin{array}{ccc}
\frac{f_{1N}+a_{1}^{0}}{\sqrt{2}} & a_{1}^{+} & K_{1}^{+} \\ 
a_{1}^{-} & \frac{f_{1N}-a_{1}^{0}}{\sqrt{2}} & K_{1}^{0} \\ 
K_{1}^{-} & {\bar{K}}_{1}^{0} & f_{1S}%
\end{array}%
\right) ^{\scalefont{0.81}\mu } \normalsize \label{LR}
\end{equation}
and $\Delta =\mathrm{diag}(\delta _{u},\delta _{u},\delta _{s})$ describes
explicit breaking of the chiral symmetry in the (axial-)vector channel. Note
that the explicit symmetry breaking in the (pseudo)scalar sector is
described by the term Tr$[H(\Phi +\Phi ^{\dagger })]$ with $H=1/2\,\mathrm{%
diag}(h_{0N},h_{0N},\sqrt{2}h_{0S})$, $h_{0N}=const.$, $h_{0S}=const.$ Also, 
$D^{\mu }\Phi =\partial ^{\mu }\Phi -ig_{1}(L^{\mu }\Phi -\Phi R^{\mu })$ $%
-ieA^{\mu }[t^{3},\Phi ]$ is the covariant derivative ($A^{\mu }$ is the
photon field); $L^{\mu \nu }=\partial ^{\mu }L^{\nu }-ieA^{\mu
}[t^{3},L^{\nu }]-(\partial ^{\nu }L^{\mu }-ieA^{\nu }[t^{3},L^{\mu }])$, $%
R^{\mu \nu }=\partial ^{\mu }R^{\nu }-ieA^{\mu }[t^{3},R^{\nu }]-(\partial
^{\nu }R^{\mu }-ieA^{\nu }[t^{3},R^{\mu }])$ are, respectively, the
left-handed and right-handed field strength tensors and the term $%
c_{1}[(\det \Phi +\det \Phi ^{\dagger })^{2}-4\det (\Phi \Phi ^{\dagger })]$
describes the $U(1)_{A}$ anomaly. Note that in this paper we are using a
different way to model the chiral anomaly (see Ref.\ \cite{Fariborz} and references therein) than in our previous papers \cite{Krakow,Madrid,Paper1,Previouspapers,Stani}
where the 't Hooft form of the chiral-anomaly term reading $c(\det \Phi
+\det \Phi ^{\dagger })$ was used \cite{Hooft}. The reason is that the
chiral-anomaly term now present in our Lagrangian (\ref{Lagrangian})
influences, as one would expect, only the phenomenology of the pseudoscalar
singlets (\ie, $\eta $ and $\eta ^{\prime }$), whereas the 't Hooft
form of the chiral-anomaly term influences the phenomenology of other mesons
(such as, \eg, the $\sigma $ states) as well \cite{Klempt}. 

In the non-strange sector, we assign the fields $\vec{\pi}$ and $\eta _{N}$
to the pion and the $SU(2)$ counterpart of the $\eta $ meson, $\eta
_{N}\equiv (\bar{u}u+\bar{d}d)/\sqrt{2}$. The fields $\omega
_{N}^{\mu }$, $\vec{\rho}^{\mu }$, $f_{1N}^{\mu }$ and $\vec{a}_{1}^{\mu }$
are assigned to the $\omega (782)$, $\rho (770)$, $f_{1}(1285)$ and $%
a_{1}(1260)$ mesons, respectively \cite{Paper1}. In the strange sector, we
assign the $K$ fields to the kaons; $\eta _{S}$ is the strange contribution
to the $\eta $ and $\eta ^{\prime }$ fields and the $\omega _{S}^{\mu } $, $%
f_{1S}^{\mu }$, $K^{\star \mu }$ and $K_{1}^{\mu }$ fields correspond to the 
$\phi (1020)$, $f_{1}(1420)$, $K^{\star }(892)$ and $K_{1}(1270)$ mesons,
respectively.

The assignment of the scalar states in our model to physical resonances is
ambiguous. In accordance with Ref.\ \cite{Paper1}, we assign the $\vec{a}_{0}
$ field to $a_{0}(1450)$ and, consequently, $K_{S}$ to the physical $%
K_{0}^{\star }(1430)$ state. This, of course, presupposes that these two
states above 1 GeV are $\bar{q}q$ states (as all the fields present in
our model are $\bar{q}q$\ states \cite{Paper1}) and thus one needs to
determine whether such an assignment allows for a global fit with a correct
description of meson phenomenology to be found.

Additionally, the Lagrangian (\ref{Lagrangian}) contains two isoscalar $J^{PC}=0^{++}$
states, $\sigma _{N}$ \{pure non-strange state, $\sigma _{N}=\bar{n}%
n\equiv (\bar{u}u+\bar{d}d)/\sqrt{2}$ \cite{Paper1}\} and $\sigma
_{S}$ (pure strange state, $\sigma _{S}\equiv \bar{s}s$).\ As noted in
Refs.\ \cite{Krakow,Madrid}, we observe mixing of $\sigma _{N}$ and $\sigma
_{S}$ in the Lagrangian leading to the emergence of two new states, $\sigma
_{1}$ (predominantly non-strange) and $\sigma _{2}$ (predominantly strange). 
The next section describes results regarding the masses and the two-pion
decay widths of the $\sigma _{1,2}$ states that allow for an assignment of
the $\sigma _{1,2}$ states to physical resonances.

In order to implement spontaneous symmetry breaking in the model, we shift $%
\sigma _{N}$ and $\sigma _{S}$ by their respective vacuum expectation values 
$\phi _{N}$\ and $\phi _{S}$. Mixing terms containing axial-vectors and
pseudoscalars and $K^{\star }$ and $K_{S}$ then arise and are removed as
described in Ref.\ \cite{Madrid}. Consequently, renormalisation coefficients
are introduced for the pseudoscalar fields and $K_{S}$ \cite{Madrid}. We
note the following formulas for $Z_\pi$ (renormalisation coefficient
of the pion) and $Z_K$ (renormalisation coefficient
of the kaon): $\phi _{N}=Z_{\pi }f_{\pi }$ \cite%
{Paper1} and analogously $\phi _{S}=Z_{K}f_{K}/\sqrt{2}$, where $f_{\pi }=92.4$ MeV and $f_{K}=155.5/\sqrt{2} $ MeV are,
respectively, pion and kaon decay constants \cite{PDG}.

The Lagrangian (\ref{Lagrangian}) contains 14 parameters: $\lambda _{1}$, $%
\lambda _{2}$, $c_{1}$, $h_{0N}$, $h_{0S}$, $h_{1}$, $h_{2}$, $h_{3}$, $m_{0}^{2}$, $%
g_{1}$, $g_{2}$, $m_{1}$, $\delta _{u}$, $\delta _{s}$. The parameter $g_{2}$
is determined from the decay width $\rho \rightarrow \pi \pi $ \cite{Paper1}%
; we set $h_{1}=0$ in accordance with large-$N_{c}$\ deliberations \cite%
{Paper1} and also $\delta _{u}=0$ because the explicit symmetry breaking is
small in the non-strange sector. All other parameters are calculated from a
global fit of masses including $m_{\pi }$, $m_{K}$, $m_{\eta }$, $m_{\eta
^{\prime }}$, $m_{\rho }$, $m_{K^{\star }}$, $m_{\omega _{S}\equiv \varphi
(1020)}$, $m_{f_{1S}\equiv f_{1S}(1420)}$, $m_{a_{1}}$, $m_{K_{1}\equiv
K_{1}(1270)}$, $m_{a_{0}\equiv a_{0}(1450)}$ and $m_{K_{S}\equiv
K_{0}^{\star }(1430)}$. We also use the full decay width of $a_{0}(1450)$ 
\cite{Paper1} in the fit to further constrain the parameters [$\Gamma
_{a_{0}(1450)}^{\mathrm{exp.}}=265$ MeV]. Note, however, that the mass terms
from the Lagrangian (\ref{Lagrangian})\ used in the fit allow only for the
linear combination $m_{0}^{2}+\lambda _{1}(\phi _{N}^{2}+\phi _{S}^{2})$
rather than the parameters $m_{0}^{2}$ and $\lambda _{1}$ by themselves to
be determined \cite{PGRKW}.

\section{The Global Fit and Two-Pion Decay of the Sigma Mesons}

\begin{table}[h] \centering%
\scalefont{0.89} 
\begin{tabular}{|c|c|c|c|c|c|c|}
\hline
Mass & $m_{\pi }$ & $m_{K}$ & $m_{\eta }$ & $m_{\eta ^{\prime }}$ & $m_{\rho
}$ & $m_{K^{\star }}$ \\ \hline
PDG Value (MeV) \cite{PDG} & 139.57 & 493.68 & 547.85 & 957.78 & 775.49 & 
891.66 \\ \hline
Our Value (MeV) & 138.65 & 497.96 & 523.30 & 957.79 & 775.49 & 916.52 \\ 
\hline
Mass & $m_{\varphi }$ & $m_{f_{1S}}$ & $m_{a_{1}}$ & $m_{K_{1}}$ & $%
m_{a_{0}} $ & $m_{K_{S}}$ \\ \hline
PDG Value (MeV) \cite{PDG} & 1019.5 & 1426.4 & 1230 & 1272 & 1474 & 1425 \\ 
\hline
Our Value (MeV) & 1036.9 & 1457 & 1219 & 1343 & 1452 & 1550 \\ \hline
\end{tabular}%
\caption{Masses from our global fit. The value of $m_{K_S}$ is larger than the PDG value due to the
pattern of explicit symmetry breaking that in our model makes strange mesons approximately 100 MeV ($\simeq$ strange-quark
mass) heavier than non-strange mesons. The relatively large value of $m_{K_1}$ is under investigation \cite{PGRKW}. 
All other values correspond very well to experimental data. Note that the fit also yields $\Gamma_{a_0(1450)} = 265$ MeV $\equiv \Gamma_{a_0(1450)}^{\rm exp.}$.}%
\label{Table1}%
\end{table}%

\begin{figure}[h]
\begin{center}
\includegraphics[
height=1.55in,
width=5.0in
]{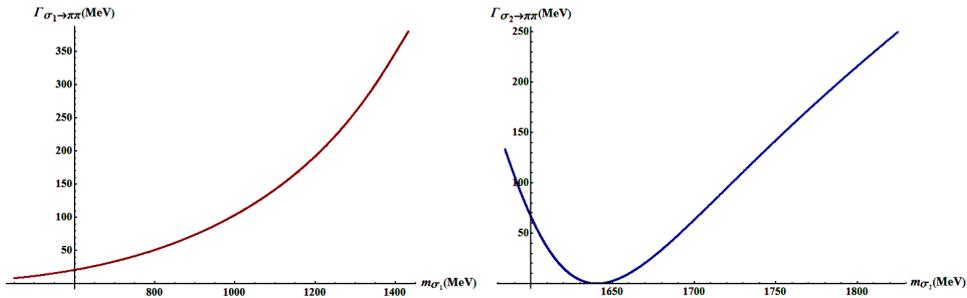}
\end{center}
\caption{$\Gamma _{\protect\sigma _{1,2}\rightarrow \protect\pi \protect\pi }
$ as function of $m_{\protect\sigma _{1,2}}$.}
\label{Figure1}
\end{figure}

Results for masses from our best fit are shown in Table \ref{Table1}.
Mixing between the pure states $\sigma _{N}$ and $\sigma _{S}$ is
implemented analogously to the quarkonium-glueball mixing of Ref.\ \cite%
{Stani}. In our case, two mixed states emerge: $\sigma _{1}$ (95\%
non-strange, 5\% strange) and $\sigma _{2}$ (95\% strange, 5\% non-strange).
Their masses and decay widths depend on seven parameters ($m_{0}^{2}$, $%
\lambda _{1}$, $\lambda _{2}$, $g_{1}$, $h_{1,2,3}$). The decay widths $%
\Gamma _{\sigma _{1,2}\rightarrow \pi \pi }$ depend on the parameters $%
m_{0}^{2}$\ and $\lambda _{1}$\ separately rather than on the linear
combination $m_{0}^{2}+\lambda _{1}(\phi _{N}^{2}+\phi _{S}^{2})$.
Therefore, using the value of the linear combination $m_{0}^{2}+\lambda
_{1}(\phi _{N}^{2}+\phi _{S}^{2})$ obtained from the fit allows us to
substitute $\lambda _{1}$ by $m_{0}^{2}$ in the formulas for $\Gamma
_{\sigma _{1,2}\rightarrow \pi \pi }$. Varying $m_{0}^{2}$ and consequently $%
m_{\sigma _{1}}$ and $m_{\sigma _{2}}$ leads to diagrams for the decay
widths shown in Fig.\ (\ref{Figure1}). (Note that $m_{0}^{2}<0$ is required
for the spontaneous symmetry breaking to be implemented correctly in the
model.) 

There is a very good correspondence of our predominantly non-strange state $%
\sigma _{1}$ with $f_{0}(1370)$ as experimental data \cite{PDG}\ regarding
the latter state read $m_{f_{0}(1370)}^{\mathrm{exp.}}=(1200-1500)$\ MeV and 
$\Gamma _{f_{0}(1370)}^{\mathrm{exp.}}=(200-500)$\ MeV with a dominant
two-pion decay channel. The state $\sigma _{2}$ corresponds well to $%
f_{0}(1710)$: experimental data regarding this resonance read $\Gamma
_{f_{0}(1710)\rightarrow \pi \pi }^{\mathrm{exp.}}=(29.28\pm 6.53)$ MeV
which leads to $m_{\sigma _{2}}^{(1)}=1613$ MeV and $m_{\sigma
_{2}}^{(2)}=1677$\ MeV; $m_{\sigma _{2}}$ is very close to $m_{f_{0}(1710)}^{%
\mathrm{exp.}}=(1720\pm 6)$ MeV. We thus conslude that both resonances $%
f_{0}(1370)$ and $f_{0}(1710)$ are predominantly quarkonia: the former 95\% $%
\bar{n}n$\ and the latter 95\% $\bar{s}s$.\ Using $m_{\sigma
_{2}}^{(1)}$ and $m_{\sigma _{2}}^{(2)}$ it is possible to calculate the
corresponding two values of $m_{0}^{2}$ and then we obtain $m_{\sigma
_{1}}^{(1)}=1360$ MeV, $\Gamma _{\sigma _{1}\rightarrow \pi \pi }^{(1)}=309$
MeV and $m_{\sigma _{1}}^{(2)}=1497$ MeV, $\Gamma _{\sigma _{1}\rightarrow
\pi \pi }^{(2)}=415$ MeV. Both sets of values are within PDG data.

\section{Summary and Outlook}

We have presented a $U(3)_{L}\times U(3)_{R}$ Linear Sigma Model with
(axial-)vector mesons. The model contains two isoscalar $J^{PC}=0^{++}$\
states: the pure $\bar{n}n$ state $\sigma _{N}$ and the pure $\bar{%
s}s$ state $\sigma _{S}$. Mixing of the pure states originates a
predominantly non-strange state $\sigma _{1}$ and a predominantly strange
state $\sigma _{2}$. In order to assign the latter states to experimentally
measured ones, we have calculated their masses and decay widths. We have
determined the model parameters by using a global fit of meson masses
(except the sigma masses) and the total decay width of $a_{0}(1450)$. Our
results regarding the masses and decay widths of $\sigma _{1,2}$ lead to
conclusion that $\sigma _{1}$ corresponds to $f_{0}(1370)$ and that $\sigma
_{2}$ corresponds to $f_{0}(1710)$. Conversely, the results imply that $%
f_{0}(1370)$ is a predominantly $\bar{n}n$ state and that $f_{0}(1710)$
is a predominantly $\bar{s}s$ state. However, one still needs to
calculate other decay widths from the $N_{f}=3$ sector and verify whether a fit
with reasonable phenomenology can be found with assumption of scalar
quarkonia in the region under 1 GeV \cite{PGRKW}.

\section{Acknowledgments}

I am grateful to Francesco Giacosa, Dirk Rischke, P\'{e}ter Kov\'{a}cs and Gy\"{o}rgy Wolf for valuable discussions regarding my work.

\end{document}